\documentclass[conference]{IEEEtran}
\IEEEoverridecommandlockouts

\usepackage{cite}
\usepackage{amsmath,amssymb,amsfonts}
\usepackage{algorithmic}
\usepackage{graphicx}
\usepackage{textcomp}
\usepackage{xcolor}
\usepackage{url}
\usepackage{tabularx}
\usepackage{booktabs} 
\usepackage{float}
\usepackage{placeins}
\usepackage{tikz}
\usetikzlibrary{positioning, shapes.geometric, arrows.meta}
\usepackage{tikz} 
\usetikzlibrary{positioning}
\usepackage{float}
\usepackage{cite}
\usepackage{tikz}
\usetikzlibrary{positioning}
\usepackage{float}
\usepackage{placeins}
\usepackage{graphicx}
\usepackage{float}
\usepackage{placeins}
\usepackage{tikz}
\usepackage[table]{xcolor}
\usepackage{colortbl}
\usepackage[T1]{fontenc}
\usepackage{lmodern}

\usepackage{tikz}
\usetikzlibrary{shapes.geometric, arrows.meta, positioning, shadows.blur, calc, backgrounds}
\usepackage{fontawesome5}
\usepackage[T1]{fontenc}
\usepackage[utf8]{inputenc}
\usepackage{lmodern} 
\definecolor{colorDataTop}{HTML}{005C97}
\definecolor{colorDataBot}{HTML}{363795}
\definecolor{colorPrepTop}{HTML}{00d2ff}
\definecolor{colorPrepBot}{HTML}{3a7bd5}
\definecolor{colorEncTop}{HTML}{4e54c8}
\definecolor{colorEncBot}{HTML}{8f94fb}
\definecolor{colorFusionTop}{HTML}{8E2DE2}
\definecolor{colorFusionBot}{HTML}{4A00E0}
\definecolor{colorUncertTop}{HTML}{f12711}
\definecolor{colorUncertBot}{HTML}{f5af19}
\definecolor{colorDecTop}{HTML}{11998e}
\definecolor{colorDecBot}{HTML}{38ef7d}

\def\BibTeX{{\rm B\kern-.05em{\sc i}\kern-.025emB\kern-.08em\TeX}}

\begin{document}

\title{An Explainable Agentic AI Framework for Uncertainty-Aware and Abstention-Enabled
Acute Ischemic Stroke Imaging Decisions}

\author{
\IEEEauthorblockN{1\textsuperscript{st} Md Rashadul Islam}
\IEEEauthorblockA{
\textit{Department of Computer Science and Engineering} \\
\textit{Daffodil International University}\\
islam15-6062@s.diu.edu.bd \\
Dhaka, Bangladesh}

}

\maketitle

\begin{abstract}
Artificial intelligence (AI) models have demonstrated considerable potential in the imaging of acute ischemic stroke, especially in the detection and segmentation of lesions via computed tomography (CT) and magnetic resonance imaging (MRI). Nevertheless, the majority of existing approaches operate as black-box predictors, providing deterministic outputs without transparency regarding predictive uncertainty or the establishment of explicit protocols for decision rejection when predictions are ambiguous. This deficiency presents considerable safety and trust issues within the context of high-stakes emergency radiology, where inaccuracies in automated decision-making could conceivably lead to negative consequences in clinical settings.\cite{Esteva2019, Topol2019}.

In this paper, we introduce an explainable agentic AI framework targeted at uncertainty-aware and abstention-based decision-making in AIS imaging. It is based on a multistage agentic pipeline.

In this framework, a perception agent performs lesion-aware image analysis, an uncertainty estimation agent estimates the predictive confidence at the slice level and a decision agent dynamically decides whether to make or withhold the prediction based on prescribed uncertainty thresholds. This approach is different from previous stroke imaging frameworks, which have primarily aimed to improve the accuracy of segmentation or classification.
\cite{Maier2017ISLES, Winzeck2018}, Our framework explicitly emphasizes clinical safety, transparency, and decision-making processes that are congruent with human values.

We validate the practicality and interpretability of our framework through qualitative and case-based examinations of typical stroke imaging scenarios.
This examination demonstrates a natural correlation between uncertainty-driven abstention and the existence of lesions, fluctuations in image quality, and the specific anatomical definition being analyzed. Furthermore, the system integrates an explanation mode, offering visual and structural justifications to bolster decision-making, thereby addressing a crucial limitation observed in existing uncertainty-aware medical imaging systems: the absence of actionable interpretability.\cite{Ghassemi2021, Holzinger2022}.

This research does not claim to establish a high-performance benchmark; instead, it presents agentic control, uncertainty-awareness, and selective abstention as essential design principles for the creation of safe and reliable MI-AI. Our results support the idea that incorporating explicit stalling behavior within agentic architectures could accelerate the development of clinically deployable AI systems for acute stroke intervention.
\end{abstract}


\section{Introduction}

Acute ischemic stroke continues to be one of the leading causes of mortality and long-term disability worldwide, placing a significant burden on patients, caregivers, and healthcare systems. Accurate and timely interpretation of medical imaging, particularly computed tomography (CT), CT angiography (CTA), and magnetic resonance imaging (MRI).Artificial intelligence (AI) has emerged as a promising tool to support radiological assessment by enabling automated lesion detection, segmentation, and triage from stroke imaging data\cite{Maier2017ISLES, Winzeck2018}.

While deep learning has advanced significantly in the domain of medical imaging, the majority of current AI systems for stroke diagnosis function as deterministic predictors, yielding consistent results without accounting for variables such as image quality, lesion ambiguity, or shifts in data distribution. This approach contrasts sharply with the practical realm of clinical radiology, where experienced radiologists often delay decision-making until additional imaging is obtained or process complex cases through multiple stages to accommodate substantial diagnostic uncertainty. In the absence of explicit mechanisms to identify images with uncertainty and abstain from decision-making, issues of safety and trust become increasingly critical, particularly in emergency situations where erroneous automated decisions could have profound clinical consequences. \cite{Esteva2019, Topol2019}.

Recent work has emphasized the importance of uncertainty quantification in medical AI and the necessity of distinguishing reliable from non-reliable predictions \cite{Abdar2021,Gal2016,Lakshminarayanan2017}. In medical imaging, uncertainty-aware approaches such as Bayesian neural networks and deep ensembles (DE) have been studied. However, they usually assume that the results can only be expressed as a numerical confidence level and would not directly apply to clincially decision making. Meanwhile, with the development of selective prediction and abstention-aware learning, models are enabled to abstain from prediction when uncertain \cite{Geifman2019,Ramos2022}. However, such methodologies are infrequently utilized in imaging analyses of ischemic stroke in a manner that reflects the systematic, staged approach of clinical evaluations.

A significant limitation of existing AI models for stroke is their lack of interpretability. Even if to date such systems have shown some interesting quantitative performances, predictions made without a clear rationale can disillusion the clinician in trusting and turning them becoming a real obstacle in clinical practice. In response to this issue, the explainable artificial intelligence (XAI) approaches have advocated saliency-based visualization and attribution techniques \cite{Selvaraju2017, Arrieta2020, Holzinger2022}. However, most existing XAI models are disconnected with uncertainty and decision-making—they merely explain prediction but fail to provide guidance on when or how a prediction is made.

Given these continuing challenges, at least one alternative holds growing appeal: that of the agentic AI approach, in which complex tasks are reduced to a set of interactively controlled subtasks or agents programmed for perception, reasoning and decision-making. Agentic and multi-stage architectures naturally provide the framework to simulate clinical workflows, where tasks like imaging interpretation, determining level of confidence and escalating a decision are immediately independent but mutually dependent. Despite the promising advance of agentic modelling overall in AI research, there has been little prior applied work on safety-critical medical imaging, such as acute stroke decision support.

We present an interpretable agentic AI approach for decision-making in acute ischemic stroke imaging that accounts for uncertain predictions and abstention. Rather than relying on a single performance measurement, e.g., predictive accuracy, our model also highlights the clinical safety and transparency issue and human value-naturalized decision making. The system consists of a modular agentic pipeline which specifically includes: a perception agent, which accomplishes lesion-aware image analysis, an uncertainty estimation agent that estimates confidence at the slice level and a decision agent that decides whether to predict or abstain using predefined thresholding when uncertainty is higher than configured by thresholds. Explainability features are added to give interpretable evidence for predictive and abstention decisions.

As a concept and exploration, the study is necessarily limited in scope from the point of view of performance. We present our own representative cases and bring attention to qualitative characteristics of the disease, in an attempt to illustrate how agentic control, uncertainty and selective abstention can all be seamlessly incorporated into AI for stroke imaging. We argue that these design principles are essential to supporting the clinical relevance of systems in high-risk emergency care settings.

\section{Related Work}
\label{sec:related_work}

\subsection{AI for Acute Ischemic Stroke Imaging}

Artificial intelligence has been thoroughly explored in the realm of acute ischemic stroke imaging, especially for detecting lesions, segmentation, and prognosis using CT, CTA, and MRI modalities. Publicly available benchmarks, such as the ISLES challenges, have been crucial in establishing evaluation methods and advancing the analysis of stroke lesions \cite{Maier2017ISLES, Winzeck2018}. 

Majority of modern methods are based on CNNs, which are typically used in multiscale or 3D manner to capture spatial context in volumetric brain imaging data \cite{Kamnitsas2017,Monteiro2020}. These approaches achieve strong quantitative results on benchmark tasks by optimizing metrics, like the Dice score or sensitivity. Yet, they frequently give high precedence of algorithmic correctness over clinical practice and persistently presume that every input should have a categorical answer even for examples that are complicated or subpar.

\subsection{Uncertainty Estimation in Medical Imaging}

Understanding uncertainty is important for trusting medical AI, especially in imaging that affects safety. Researchers are studying methods like Bayesian neural networks, Monte Carlo dropout, and ensemble techniques to measure two types of uncertainty in deep learning models. \cite{Gal2016, Lakshminarayanan2017}. In medical imaging, using methods that consider uncertainty can help find predictions that might not be reliable and cases that are unclear \cite{Abdar2021}.

Even with these improvements, uncertainty in stroke imaging is often just shown as an extra confidence score. In usual procedures, these uncertainty estimates do not change how the system works. Systems do not change their decisions based on these estimates or send uncertain cases to experts for review.

\subsection{Selective Prediction and the Limits of Monolithic Models}

Selective prediction frameworks can avoid making predictions when they are uncertain, which helps to increase the reliability of the results\cite{Geifman2019}. In healthcare, using this method can lower the chances of making harmful mistakes when the model is too confident\cite{Ramos2022}.

Abstention is common in medical AI, once a model has made its decision. It does not sufficiently distinguish between perception, acknowledging uncertainty, and decision making. Because of this, the reasons for not making a prediction are often unclear and may not fit well with real clinical practice. This is a problem in urgent situations like stroke imaging. In emergency radiology, it is important that decisions are reliable and safe, as well as accurate.

\vspace{0.5em}

\begin{table*}[t]
\centering
\caption{Comparative synthesis of representative AI-based approaches for acute ischemic stroke imaging, emphasizing recent advances (2024--2025) and highlighting gaps in uncertainty-aware abstention and agentic decision control.}
\label{tab:comparison_stroke_sota}

\scriptsize
\renewcommand{\arraystretch}{1.35}
\setlength{\tabcolsep}{3.6pt}

\begin{tabularx}{\textwidth}{|
>{\centering\arraybackslash}p{1.2cm}|
>{\centering\arraybackslash}p{1.0cm}|
>{\centering\arraybackslash}p{2.6cm}|
>{\centering\arraybackslash}p{1.8cm}|
>{\centering\arraybackslash}p{1.7cm}|
>{\centering\arraybackslash}p{1.5cm}|
>{\centering\arraybackslash}p{1.5cm}|
X|
}
\hline
\textbf{Ref.} &
\textbf{Year} &
\textbf{Core Method} &
\textbf{Clinical Focus} &
\textbf{Uncertainty} &
\textbf{XAI} &
\textbf{Agentic} &
\textbf{Key Limitation / Gap} \\
\hline

\cite{Maier2017ISLES} &
2017 &
CNN Ensemble &
Stroke MRI Segmentation &
No &
No &
No &
Optimizes segmentation accuracy only; lacks uncertainty modeling and decision control. \\
\hline

\cite{Winzeck2018} &
2018 &
Hybrid ML + CNN &
Stroke Outcome Prediction &
No &
Partial &
No &
Deterministic predictions; limited transparency and no abstention strategy. \\
\hline

\cite{Abdar2021} &
2021 &
Uncertainty-aware DL (Survey) &
Medical Imaging (General) &
Yes &
Partial &
No &
Methodological overview; no task-specific integration for stroke workflows. \\
\hline

\cite{Ramos2022} &
2022 &
Abstention-aware CNN &
Medical Diagnosis &
Partial &
No &
No &
Introduces rejection option but lacks explainability and clinical workflow alignment. \\
\hline

\cite{Zou2024} &
2024 &
Uncertainty Modeling Survey &
Medical AI (General) &
Yes &
Partial &
No &
Focuses on uncertainty theory; does not address agentic decision-making in stroke. \\
\hline

\cite{delaRosa2025} &
2025 &
DeepISLES (CNN) &
Stroke MRI Segmentation &
No &
No &
No &
Strong benchmark performance; no safety-aware abstention or agentic reasoning. \\
\hline

\cite{Tzanis2025} &
2025 &
mAIstro (Multi-Agent System) &
Radiomics &
No &
Yes &
Yes &
Agentic design without explicit uncertainty-driven abstention in acute stroke settings. \\
\hline

\cite{Feng2025} &
2025 &
M\textsuperscript{3}Builder &
Medical Imaging Models &
Partial &
Yes &
Yes &
Focuses on model construction; lacks real-time clinical decision control. \\
\hline

\textbf{This Work}\textsuperscript{$\dagger$} &
\textbf{2026} &
\textbf{Explainable Agentic Framework} &
\textbf{Acute Stroke CT/CTA} &
\textbf{Yes} &
\textbf{Yes} &
\textbf{Yes} &
\textbf{Conceptual integration of uncertainty-aware abstention, explainability, and agentic decision control for safety-critical stroke imaging.} \\
\hline

\end{tabularx}

\vspace{0.5em}
\footnotesize{\textsuperscript{$\dagger$}Conceptual and exploratory framework positioned for next-generation clinically deployable stroke imaging AI systems.}
\end{table*}

\subsection{Explainable Artificial Intelligence in Clinical Imaging}

Explainable artificial intelligence (XAI) is considered essential for improving transparency, accountability, and clinician confidence in medical AI systems.Commonly, visualization approaches such as Grad-CAM are utilized to visualize which parts of the image have impact on model decisions \cite{Selvaraju2017}. Surveys have also overall stressed the importance of explainability of for regulatory compliance, ethical deployment, and clinical adoption \cite{Arrieta2020, Holzinger2022}.

However, most of these explainability methods involve a post-hoc processing step, which is usually separate from guidance based on uncertainty and the decision-making process. Therefore, explanations often help us understand how a decision was made, rather than addressing the more important question of whether the decision was appropriate. This is especially important in clinical settings, where the consequences can be very serious.

\subsection{Agentic and Multi-Stage AI Systems in Healthcare}

Agentic and multi-staged AI systems, in this context, disassemble processes into mediated components, a structure that underpins sensing (which provides function), decision-making, and action (which requires such function). These architectures have recently gained attention in clinical decision support, mainly because of their modular and interpretable nature \cite{Shortliffe2021,Kiani2020}.

Recent studies using multiple agents in medical imaging show the potential for training structured collective intelligence \cite{Tzanis2025, Feng2025}. However, these systems have not yet been used in the acute stroke imaging process, nor have they integrated uncertainty-based abstention as a key safety feature in their design.

\subsection{Summary and Gap Analysis}

State-of-the-art in stroke imaging AI, uncertainty estimation, abstention-aware learning, explainable AI and agent systems represent significant recent advances in these areas. However, these two regions have evolved independently. Existing approaches are unable to find a tradeoff between uncertainty-aware abstention, explanation and Reactive Decision Control (RDC) with the certainty, and they fail to seamlessly combine all these aspects into an efficient solution for AIS imaging.

This study aims to address this gap by presenting an explainable agentic AI architecture. This architecture is specifically designed to integrate lesion-aware perception, uncertainty estimation, selective abstention, and clinician-aligned decision support, with the goal of improving the safety and reliability of stroke imaging systems.

\section{Methodology}
\label{sec:methodology}

This section delineates the Explainable Agentic AI framework, which is pertinent to uncertainty-aware and abstention-enabled acute ischemic stroke imaging. This approach is intentionally similar to clinical reasoning, where perception, changes in beliefs that aren't fully formed, and the process of making decisions are seen as interacting functional components. The model's foundation goes beyond just predictions; it also considers safety, how easy it is to understand, and how well it matches medical knowledge.

\subsection{Overview of the Agentic Framework}

This work introduces a modular architecture for an agent-based system, which (incorporating various specialized agents) now offers not only secure and interpretable decision-support in the field of stroke-imaging. Unlike monolithic deep learning approaches that directly fuse images and predictions, the proposed approach decouples perception from uncertainty estimation and decision. The structure of this model mirrors the way in which radiologists read clinically, considering speed of image reading, confidence assessment and how to handle cases which are not certain together but found them as separate components.

\begin{figure}[!ht]
\centering
\resizebox{0.9\columnwidth}{!}{
\begin{tikzpicture}[
    node distance=1.4cm,
    every node/.style={font=\sffamily},
    premium3d/.style={
        rectangle, rounded corners=6pt,
        minimum width=7cm, minimum height=1.55cm,
        draw=gray!25, fill=white,
        blur shadow={shadow blur steps=8, shadow xshift=3pt, shadow yshift=-3pt},
        align=center, inner sep=10pt, line width=0.9pt
    },
    thick_arrow/.style={-Stealth, line width=2.6pt, color=gray!35, rounded corners=10pt},
    side_tag/.style={font=\sffamily\bfseries\tiny, color=gray!65, align=right, rotate=90}
]

\node (input) [premium3d, top color=gray!5, bottom color=gray!15] {
    \textbf{\large \faDatabase\ Acute Stroke Imaging Input}\\
    \scriptsize CT / CTA / MRI Brain Slices $\cdot$ Emergency Imaging\\
    \textit{\tiny \faClock\ Time-critical acquisition: ``Time is Brain''}
};

\node (perception) [premium3d, below=of input,
top color=blue!10, bottom color=blue!22, draw=blue!35] {
    \textbf{\large \faEye\ Perception Agent}\\
    \textit{\scriptsize Lesion-Aware Image Representation}\\
    \rule{5cm}{0.4pt}\\
    \footnotesize Feature Extraction $\cdot$ Lesion Localization $\cdot$ Spatial Learning
};

\node (uncertainty) [premium3d, below=of perception,
top color=purple!10, bottom color=purple!22, draw=purple!35] {
    \textbf{\large \faChartArea\ Uncertainty Estimation Agent}\\
    \textit{\scriptsize Confidence \& Reliability Assessment}\\
    \rule{5cm}{0.4pt}\\
    \footnotesize Epistemic Uncertainty $\cdot$ Slice-level Confidence $\cdot$ Ambiguity Quantification
};

\node (decision) [premium3d, below=of uncertainty,
top color=orange!10, bottom color=orange!22,
draw=orange!50, line width=1.6pt] {
    \textbf{\large \faBalanceScale\ Decision Agent (Abstention-Enabled)}\\
    \vspace{0.1cm}
    \tikz[baseline=-0.5ex]{
        \node[diamond, draw=orange!70, fill=white,
        inner sep=1.2pt, font=\tiny\bfseries]
        {Uncertainty $>$ $\tau$ (Safety Threshold)};
    }\\
    \footnotesize
    \textcolor{green!60!black}{\faCheckCircle\ Predict}
    \quad | \quad
    \textcolor{red!70}{\faPauseCircle\ Abstain (Refer to Clinician)}
};

\node (output) [premium3d, below=of decision,
top color=teal!10, bottom color=teal!22, draw=teal!35] {
    \textbf{\large \faFileMedical\ Explainable Clinical Output}\\
    \scriptsize \textbf{Explainability:} Saliency Maps $\cdot$ Decision Rationale\\
    \textit{\tiny \faSearchPlus\ Clinician-Interpretable Evidence \& Transparency Module}
};

\draw [thick_arrow] (input) -- (perception);
\draw [thick_arrow] (perception) -- (uncertainty);
\draw [thick_arrow] (uncertainty) -- (decision);
\draw [thick_arrow] (decision) -- (output);

\draw [thick_arrow, dashed, color=gray!25]
(decision.east) -- ++(1.5,0) |- 
node[pos=0.25, side_tag, xshift=2.6cm]
{CLOSED-LOOP SAFETY OVERSIGHT}
(input.east);

\node [side_tag, left=of perception, xshift=1.2cm] {REPRESENTATION};
\node [side_tag, left=of uncertainty, xshift=1.2cm] {ANALYTICS};
\node [side_tag, left=of decision, xshift=1.2cm, color=orange!85] {SAFETY LOGIC};

\end{tikzpicture}
}
\caption{The proposed explainable agentic AI framework for uncertainty-aware acute stroke imaging employs a hierarchical top-down design. This design distinctly separates perception, uncertainty estimation, and safety-aware decision-making processes. Such a structure facilitates abstention in situations of high epistemic uncertainty and allows for clinician-in-the-loop intervention.}
\label{fig:agentic_framework}
\end{figure}
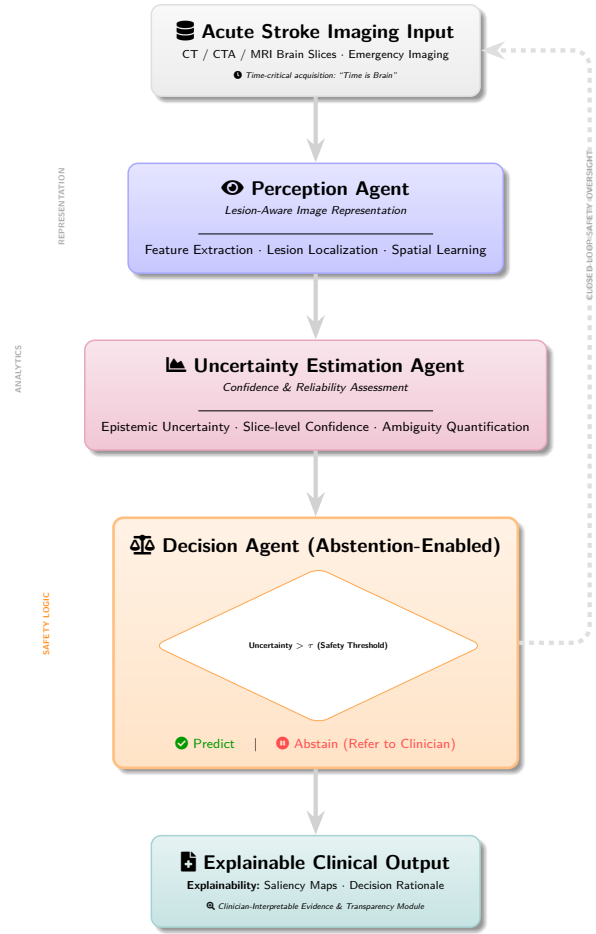

Figure~\ref{fig:agentic_framework} The system's design is distinguished by a top-down hierarchical structure. Acute stroke imaging data serve as input for the perception agent, whose outputs are then evaluated by an uncertainty estimation agent. Following this, the decision-making agent uses a safety policy that considers the possibility of not acting. At the same time, an explainability module provides clear clinical results.

\subsection{Perception Agent: Lesion-Aware Image Analysis}

The perceptual agent is designed to identify features related to lesions in imaging studies of acute stroke, including CT, CTA, and MRI scans. Functioning as a deep learning–based feature extractor, this agent is trained to prioritize relevant brain regions indicative of ischemic damage.

It is essential to understand that the perception agent does not make clinical determinations. Rather, it extracts intermediate feature representations that summarize spatial lesion attributes, contextual image data, and structural markers. This distinction ensures that perceptual analysis is separate from subsequent decision thresholds, thereby enabling uncertainty reasoning and abstention strategies to be grounded in interpretable intermediate levels, rather than an inscrutable final prediction.

\subsection{Uncertainty Estimation Agent}

The uncertainty estimation engine evaluates the validity of perceptual representations generated by the perception engine. Additionally, uncertainty is computed at the slice level to capture local ambiguities arising from factors such as low contrast, motion blur, poor lesion visibility, or complex anatomical structures.

The agent produces a normalized uncertainty score that reflects prediction confidence rather than raw class probability. This score is based on the estimation of epistemic uncertainty, indicating areas where the model's uncertainty is significant enough to preclude reliance on automated decision-making.

\begin{figure}[t]
    \centering
    \includegraphics[width=0.9\linewidth]{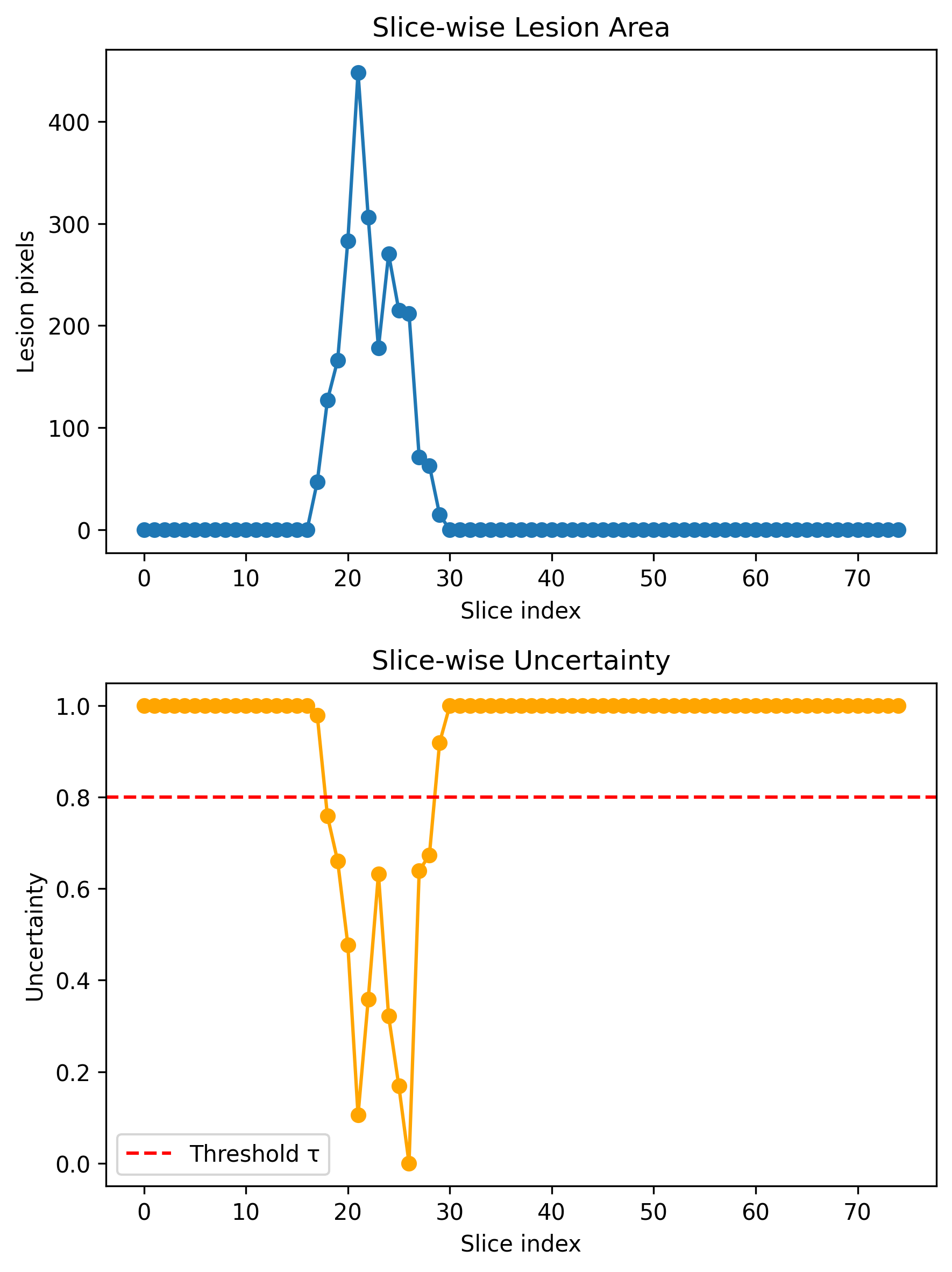}
    \caption{Slice-wise uncertainty profile across axial brain slices. Elevated uncertainty is observed in ambiguous or low-information regions, motivating abstention in safety-critical decision-making.}
    \label{fig:uncertainty_curve}
\end{figure}

Figure~\ref{fig:uncertainty_curve} presents a representative slice-wise uncertainty profile.Uncertainty is expected to grow in areas with blurred border appearances or insufficient diagnostic matter. This observation is an argument for using abstention in the case of safety-critical applications.

\subsection{Decision Agent and Abstention Mechanism}

The choice element assesses perceptual inputs along with their associated uncertainty estimates to determine if there is enough evidence for the network to make a decision. A prediction is made only when the uncertainty score falls below a specified safety threshold, represented as $\tau$. If the uncertainty surpasses this threshold, the system refrains from making a judgment, similar to a clinician who seeks further expert consultation or diagnostic testing.

This 'abstention' mechanism transforms ignorance from a passive diagnostic state into an active decision signal. By including both deferral and uncertainty, we implement a "safety-first" approach. This method discourages overconfident decision-making, which in turn reduces the chance of overly confident predictions in situations where the diagnosis is unclear.

\subsection{Explainability and Decision Transparency}

Explainability metrics are incorporated to produce human interpretable explanations for both predictions and non-decisions. The visual attribte methods showcase the predilections of the images and signal, which produce lesion-aware representations and the uncertainty-informed pathways, explaining where abstinence is advised.

\begin{figure}[t]
    \centering
    \includegraphics[width=0.95\linewidth]{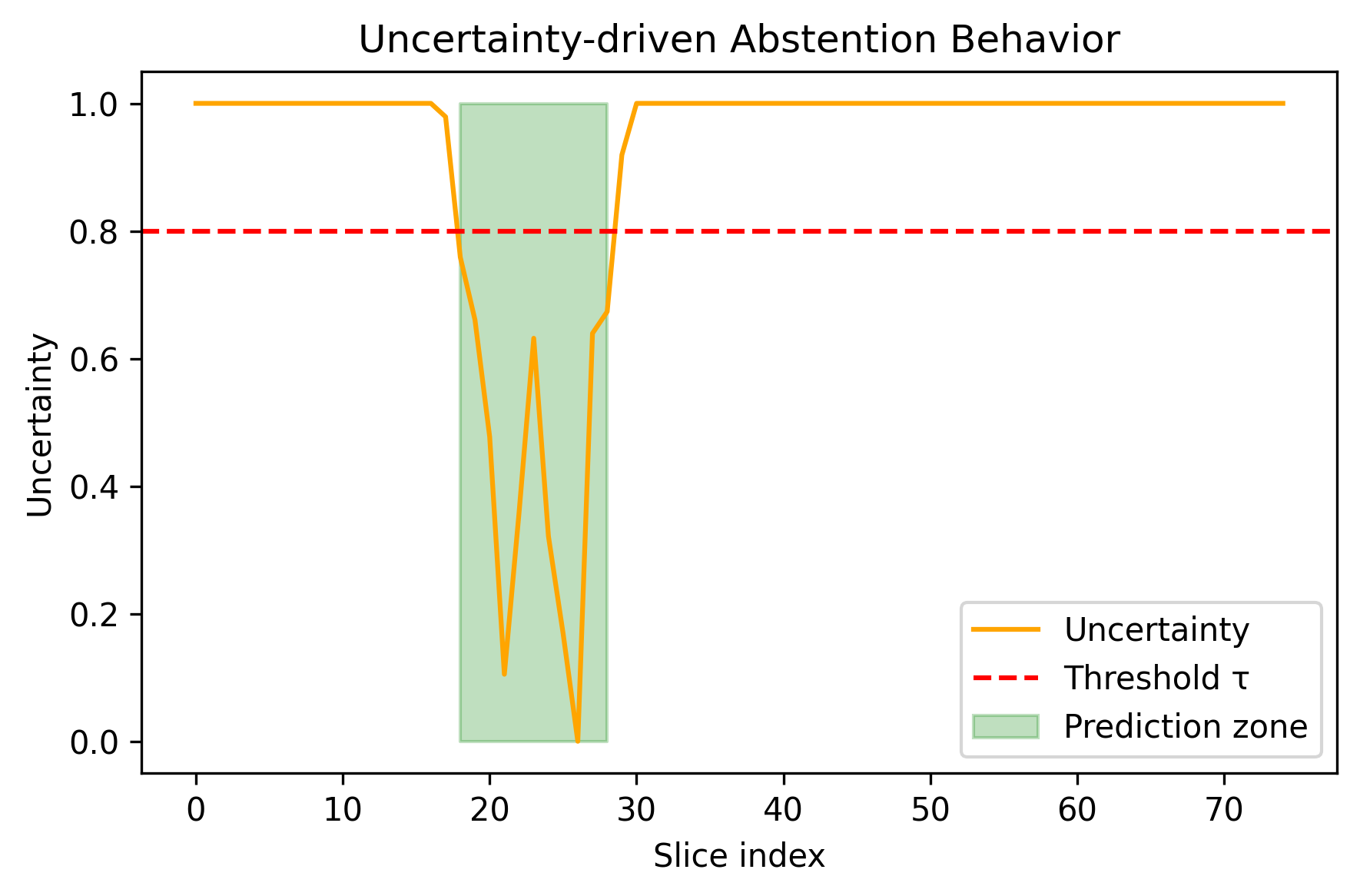}
    \caption{The visualization of decision outcomes influenced by uncertainty across representative image slices. Slices characterized by low uncertainty result in confident predictions, whereas those with high uncertainty necessitate abstention and referral to a clinician.}
    \label{fig:abstention_single}
\end{figure}

Figure~\ref{fig:abstention_single} Samples illustrate the difference of predicting and not-predicting outcomes within individual image slices. Sections with clearly defined lesion characteristics are associated with high confidence decisions, meanwhile abstention takes place in regions of uncertainty and misdiagnosis.

\vspace{0.3cm}

\begin{figure}[t]
    \centering
    \includegraphics[width=\linewidth]{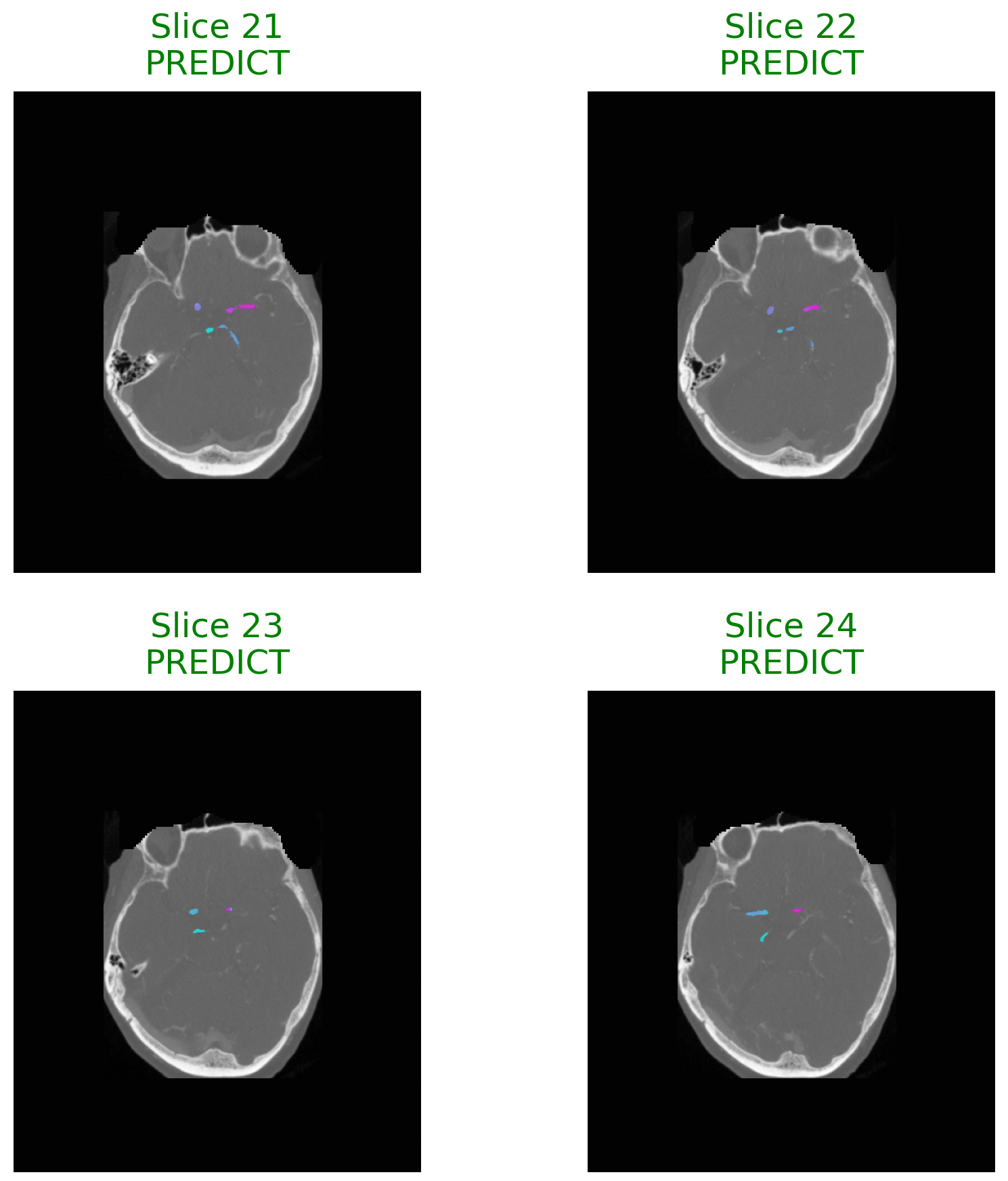}
    \caption{visualizing decision outcomes influenced by uncertainty on axial brain slices, we gain a deeper understanding of system behavior. In diagnostically clear sections, confident predictions are made, whereas in adjacent or ambiguous areas, abstention occurs, mirroring clinically appropriate deferral behavior.}
    \label{fig:abstention_grid}
\end{figure}

\subsection{Design Rationale and Clinical Alignment}

The agentic model outlined in this paper highly values clinical safety, interpretability and transparency through the avoidance of full automation. The system emulates a clinician's thought and decision explanation patterns by employing methods that delay actions in the precence of ambiguity and also incorporating explainable portions

This framework is not a stand-alone diagnostic aide but should be used as a decision support system to aid the clinician in formulating considered judgements in the milieu of acute stroke care.The design's emphasis on safety in stroke detection stems from the critical need to identify instances where predictions should be suppressed, a requirement as essential as generating precise predictions. 

Furthermore, the modular separation of perception, uncertainty estimation, and decision control promotes structured reasoning concerning system behavior and provides a basis for subsequent enhancements toward more sophisticated clinical decision-support systems.

\section{Results}
\label{sec:results}

This section presents a qualitative and behavioral evaluation of the proposed agentic AI concept. The findings not only corroborate traditional performance metrics but also offer insights into the system's responses to uncertainty, its transitions into and out of multirobot states (abstention dynamics), and its adherence to sensitivity standards required at a clinical safety level. This evaluation paradigm was deliberately chosen to reflect the exploratory and safety-critical nature of acute stroke imaging, where understanding \emph{when} and \emph{why} a model defers is as crucial as its predictive performance.

\subsection{Slice-wise Lesion Awareness and Uncertainty Behavior}

The perception agent consistently identifies lesion-associated regions across successive image slices, thus producing spatially coherent intermediate representations. Nevertheless, the system does not assign uniform reliability to all slices. Rather, it permits uncertainty to fluctuate dynamically, contingent upon both the visibility contrasts of the lesion and the surrounding anatomical context.

As depicted in Figure~\ref{fig:uncertainty_curve}, uncertainty persists at a high level within slices exhibiting limited lesion information or ambiguous visual patterns, whereas it diminishes in slices where lesion characteristics are more distinctly apparent.
The observed behavior suggests that the uncertainty estimation agent prioritizes substantial image content, rather than producing uniform confidence values throughout the volume.

The slices toward the boundary of a lesion often have intermediate levels of uncertainty, which corresponds to partial visibility of the lesions. This conclusion is consistent with current clinical knowledge, since these slices are often associated with high interpretative complexity even for an experienced radiologist.

\subsection{Uncertainty-Driven Abstention Patterns}

The policy explicitly incorporates abstentions based on uncertainty estimates. When uncertainty surpasses the predetermined safety threshold, the system refrains from processing and instead defers; it does not terminate the process to avoid potentially excessive predictions.

Figure~\ref{fig:abstention_grid}
The summary of the text encompasses two predictions and abstentions concerning axial slices. The model's interpretability indicates that confident predictions are made on slices exhibiting localized lesion structures, whereas abstentions predominantly occur in slices with low contrast, limited diagnostic cues, or ambiguous anatomical locations. 

The concept of selective deferral implies that the system is not just shifting from abstinence, but is actively making choices to defer in bioequivalently undetermined cases. This approach minimizes the risk of false positives or false negatives becoming overly confident in safety-critical situations.

\subsection{Clinical Interpretability of Decision Outcomes}

In addition to providing a binary prediction or refusal, the model offers an interpretable visualization that enhances the clinical relevance of its decision. Lesion-aware representations, which augment the input volume, are highlighted through saliency-based visualizations, while uncertainty-informed decision logic offers explanations when abstention is triggered in specific slices. 

 From a clinical perspective, it is transparent to clinicians when the model is confident and legitimately supportive, and when additional expert review may be necessary. The technology does not obscure but rather reveals uncertainty as part of the decision-making process.

\subsection{Safety-Oriented System Behavior}

The emergent behavior of the set-theoretical model, which is conservative and security oriented, has important consequences. In contrast to most traditional end-to-end systems which are designed to make predictions for all the inputs, after some exploration, the agentic network tends not to act upon uncertainty.

This feature would be particularly important considering, for instance, acute stroke care, where false-positive automatic decisions could have deleterious and irreversible consequences. Uncertainty and deferral often feature in human medical reasoning, as does the decision theory focus on these aspects. Relations also to Fischer's argument are explored.

\subsection{Summary of Observed System Properties}

The qualitative data suggests that the proposed agentic framework:
\begin{itemize}
    \item creates lesion-aware representations that preserve spatial consistency throughout image slices.
    \item The data exhibit substantial variations in slice-level uncertainty, which are associated with the visibility of lesions.
    \item selectively causes abstention in areas that are clinically unclear.
    \item articulates predictions and abstentions in a manner that is comprehensible.
    \item Demonstrates a prudent, safety-oriented approach to decision-making that aligns with established clinical protocols.
\end{itemize}
The findings demonstrate the effectiveness of agentic, uncertainty-aware decision support systems in acute stroke imaging, especially when the focus is on interpretability and safety rather than precise prediction.

\section{Discussion}
\label{sec:discussion}

In the context of acute ischemic stroke imaging, we propose an explainable agentic AI system designed for uncertainty-aware decision support. Our model has been crafted with safety and generalization of the decision in mind, thus enhancing safety compliance, interpretability and alignment with established clinical decision-making processes. This approach is in contrast to the common end-to-end neural network-based methods which make predictions without an explicit estimation of uncertainty. The basis for this design is grounded in a growing body of empirical evidence, which has shown the hazards of overreliance on “overconfident” model behavior within high-stakes applications, including in medical AI tasks where mispredictions could cause significant clinical harm\cite{Zou2024,Abdar2021}.

One of the key contributions in this work is that it provides a model to treat uncertainty as an actionable decision rather than simply as measure of confidence. It has been shown in related studies that many medical imaging systems provide uncertainty scores/confidence values which have no impact on the system's confidence to make a prediction, despite exhibiting large epistemic uncertainty \cite{Ramos2022,Zou2024}. On the other hand, when the degree of uncertainty exceeds a user-defined safety threshold, our decision agent design enables us to rely on an abstention mechanism and we present principles toward its development.This approach aligns more closely with real-world clinical practice, where cases involving slice-level assessments are referred back for expert review rather than being compelled to reach a resolution.

The qualitative findings demonstrate that the network avoids certain regions, and it is not distributed equally over all image slices, but instead clustered within those regions that are diagnostically more challenging due to a lack of structural information (e.g., low-contrast tissue), unclear lesion boundaries or complex anatomy. Recent issues in stroke imaging have shown how this segmentation model's performance is not optimal at the boundaries between lesions and in ambiguous tissue contrast \cite{delaRosa2025}. As a result, instead of random rejection or nonsensical output, the behaviors of abstention patterns indicate that the uncertainty estimator learns to governclinically significant ambiguity properly.

Moreover, enhancing explainability improves the clinician-friendliness of our approach. Prior research has demonstrated that black-box predictions can engender clinician skepticism, which subsequently affects regulatory approval and the secure implementation of medical imaging systems~\cite{Abdar2021,Feng2025}. The framework's saliency-based visual explanations of both predictive and abstain decisions enable clinicians to discern the image regions prioritized by the model and the rationale behind any decision-making delays. Interpretability is, in essence, a fundamental aspect of transparency within safety-critical decision-making, rather than a secondary consideration.

Furthermore, the agentic design paradigm fundamentally alters how AI is used in acute stroke workflows. The system is designed to support physicians' decisions, rather than acting as an independent diagnostic tool. This approach aligns with current trends in medical AI, which emphasize human involvement or agentic architectures for managing complex clinical situations, where predictive accuracy depends on interpretability, accountability, and reliability \cite{Tzanis2025,Feng2025}. Our proposed framework offers a clear method for safely integrating AI into acute stroke care by separating perception and uncertainty reasoning from the decision-making process.

\section{Limitations}
\label{sec:limitations}

Several limitations of this study warrant acknowledgment. Firstly, we have intentionally opted for a qualitative and exploratory analysis at this stage, focusing on behavioral patterns, uncertainty dynamics, and abstention profiles rather than quantitative benchmarks. While this approach aligns with recent calls for a safety-centric evaluation of medical AI systems \cite{Zou2024}, we recommend that future research incorporate large-scale quantitative measures alongside the insights presented here.

Secondly, the study is based on a limited number of sample cases. This provides a detailed insight into each individual case, but does not represent the entire patho-anatomic heterogeneity which has been reported in multi-center cohorts of stroke patients. Earlier work on the AI applied to stroke imaging has shown that, as in this case, proving generalizing by translational testing is still pending requires validation \cite{delaRosa2025}.

Thirdly, the uncertainty estimates employed in this framework measure epistemic uncertainty rather than providing a formally calibrated probabilistic guarantee. Although the uncertainty behavior aligns with clinical intuition, careful calibration and comparison with expert annotations are crucial future directions \cite{Abdar2021}.

The framework is not meant to be used as a standalone clinical decision support system. It is structured in accordance with modern regulations and ethics with regards to data, and it will serve as support rather than a replace the role of the clinician.Clinical validation and integration into the clinical workflow will be conducted in prospective studies prior to real-world application.

\section{Conclusion}
\label{sec:conclusion}

This study presents a new artificial intelligence model designed to help with decision-making in stroke imaging. The model uses a combination of different parts that handle perception, uncertainty, and the decision-making process.

Unlike earlier models that only made predictions, this system is designed to be safer and more transparent. A key feature of this system is its ability to avoid making decisions when it isn't sure. This is similar to how doctors often ask for more information when faced with a difficult medical case.

The system seems to be good at identifying hard image areas, which suggests that it understands important uncertainty types. This AI model is meant to help medical professionals, not replace them, as long as it caters caution.

It reflects the current paradigm of medical AI as it end-weeks human involvement over interpretability and risk awareness. Therefore, this study suggests that such AI can increase the safety and reliability of medical image. The model serves as the foundation to create AI that are able handle uncertainty and ensure decision security when working with high-risk clinical situations.

\bibliographystyle{IEEEtran}
\bibliography{references}

\end{document}